\documentclass[prd,showpacs,showkeys,aps,twocolumn]{revtex4}
\usepackage{graphicx}
\usepackage{epsfig}

\newcommand{\bea}{\begin{eqnarray}}
\newcommand{\eea}{\end{eqnarray}}

\newcommand{\p}{{\bf p}}
\renewcommand{\k}{{\bf k}}
\def\be{\begin{equation}}
\def\ee{\end{equation}}

\def\removetext#1{{}}

\begin{document}

\title{Universal threshold enhancement}

\author{A. Patk{\'o}s}
\email{patkos@ludens.elte.hu}
\affiliation{Department of Atomic Physics, E{\"o}tv{\"o}s University,
 H-1117 Budapest, Hungary}

\author{Zs. Sz{\'e}p}
\email{szepzs@achilles.elte.hu}
\affiliation{Research Group for Statistical Physics of the Hungarian 
Academy of Sciences, H-1117, Budapest, Hungary}

\author{P. Sz{\'e}pfalusy}
\email{psz@galahad.elte.hu}
\affiliation{
Department of Physics of Complex Systems,E{\"o}tv{\"o}s University,
H-1117 Budapest, Hungary\\
and \hbox{Research Institute for Solid State Physics and Optics,
Hungarian Academy of Sciences, H-1525 Budapest, Hungary}}

\begin{abstract}
By assuming certain analytic properties of the propagator, it is shown that
universal features of the spectral function including threshold enhancement
arise if a pole describing a particle at high temperature approaches in the
complex energy plane the threshold position of its two-body decay with the
variation of $T$.  The case is considered, when one can disregard any other
decay processes. The quality of the proposed description is demonstrated
by comparing it with the detailed large $N$ solution of the linear $\sigma$
model around the pole-threshold coincidence.
\end{abstract}

\maketitle

The near-threshold two-body decay of a resonance 
 is a well-studied problem in scattering theory 
\cite{walecka}. In a medium with
tunable parameters (e.g. temperature, density etc.) new features may 
arise \cite{yoko}. It has been found \cite{hatsuda,kuni} that a hallmark of 
partial
chiral symmetry restoration in a medium is the enhancement of the
scalar-isoscalar spectral function near the two-pion threshold. In order
to elucidate further the origins and the nature of this phenomenon, in 
this note we formulate a general approach to it. On the basis of this
investigation some features of threshold enhancement appear to be 
{\it universal}, showing up whenever the mass of a stable
particle can be driven by 
tuning some parameters of its environment to pass the threshold of a
two-body decay into stable final particles. Following our 
general discussion we shall show evidence for this situation in the
linear $\sigma$ model by studying  the propagator of the $\sigma$ 
particle, neglecting the finite in-medium lifetime of the pions.

Let us start our general discussion by assuming the existence of a branch
point in the complex $\sqrt{p^2}$ 
rest frame energy plane on the real energy axis for the propagator 
$G_\sigma(p^2,|\p|),~(p^2=p_0^2-\p^2)$, which corresponds to the 
$\sigma\rightarrow \alpha_1+\alpha_2$ hypothetical decay. 
(In this part of our discussion $\sigma$ refers to an arbitrary
particle. The extra dependence on the 3-momentum is a medium
effect.)  We can restrict
the discussion to the halfplane ${\rm Re}\sqrt{p^2}>0$, due to the 
symmetric
behavior of the propagator. At the branch point  a square root
singularity, proportional to $\delta\equiv\sqrt{1-p^2/M^2}$,
shows up in the self-energy contribution to the propagator,
 where $M$ denotes the temperature dependent position of the branch point.
The analytic (Riemann-sheet) structure of the $\sqrt{p^2}$-plane is 
specified by 
requiring that $G_\sigma$ is analytically continued through the part of 
the real axis where $p^2>M^2$, i.e. where the complex threshold factor $\delta$
is purely imaginary. Furthermore the plane we are working on is bounded by 
a cut along the section $p^2<M^2$ of the real axis. Along the cut 
$\delta$ is real positive (negative)
on the upper (lower) halfplane. Conventionally, one associates the upper
halfplane with the first, the lower one with the second Riemann-sheet.
We shall call the upper edge of the real axis reached from the first
Riemann sheet as physical, and the lower edge of the cut as unphysical.

The goal of our analysis is to investigate the threshold
behavior of the spectral function in the neighborhood of the branch
point in a small interval around some temperature $T^*$, 
to be specified below. We first restrict the discussion to the case
$|\p|=0$, which makes our analysis easier to follow. 
There are now two small quantities characterizing this situation:
$\epsilon\equiv (T-T^*)/T^*$ and $\delta$. [It is important to note,
that it is $M(T)$ and not $ M(T^*)$, which appears
in the definition of $\delta$.] Our basic assumption is that 
the inverse propagator $G_\sigma^{-1}(p_0)$ can be expanded in terms of
these two independent  variables. Then to quadratic order one 
obtains
\bea
-G_\sigma^{-1}(p_0)&\approx& a_1\epsilon+\tilde{a}_2\epsilon^2+ b_1\delta
+b_2\delta^2+c\delta\epsilon\nonumber\\
&\approx&
\frac{1}{1+ c\epsilon/ b_1}
[a_1\epsilon+a_2\epsilon^2+ 
b_1\delta+b_2\delta^2],
\label{expand}
\eea
where $a_2=\tilde a_2+a_1c/b_1$ was defined, after factoring out
a temperature dependent finite constant.

We investigate the behavior of the spectral function, which can be
derived from (\ref{expand}) by studying the pole structure starting at
high temperature. In the concrete case of the $\sigma$-model the tendency to
restore the chiral symmetry with increasing temperature reduces the
$\sigma$-mass below $2m_\pi(T).$ For such temperature $\sigma$ is
necessarily stable if the pions are assumed to be stable. 


When decreasing the temperature the
location of the $\sigma$-pole moves along the physical real axis to
the two-body decay threshold from below. Let us assume that this pole 
reaches the branch point at $T=T^*$. This is 
the temperature $T^*$ around which
the behavior of the spectral function will be investigated.

\begin{figure}[htpb]
\begin{center}
\includegraphics[width=6.5cm]{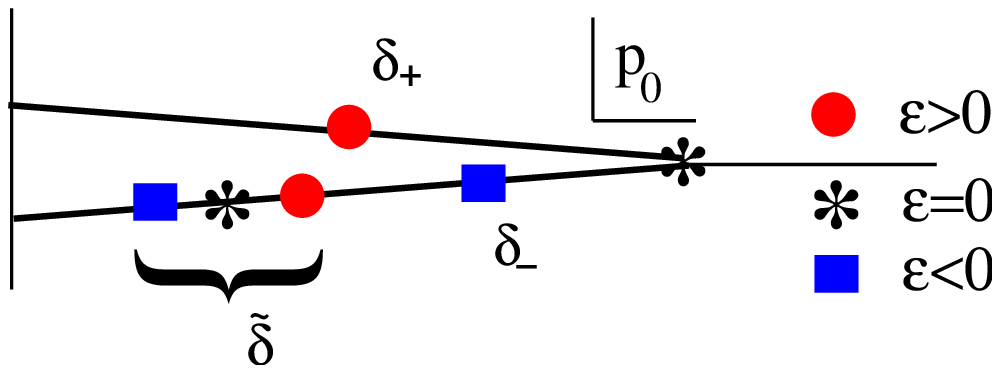}
\end{center}
\vspace*{-0.6cm}
\caption{Schematic presentation of the pole evolution.}
\label{Fig:edge}
\end{figure}

The stability of the $O(N)$ multiplet enforced by the
chiral symmetry is now postulated, without reference to any 
particular symmetry, for the general discussion. We
assume that for $\epsilon >0$ in addition to the approximate 
stability of the particles $\alpha_1$ and  $\alpha_2$, $\sigma$ is stable.
This requires a single pole, $\delta_+$, for $G_\sigma$ on the physical real
axis below the threshold $(1>\delta>0)$, corresponding to a stable
state.  
The requirement, that no other positive root should be present
is fulfilled only if $a_1<0, b_1>0, b_2>0$. 
This choice implies the existence of a root $\tilde\delta<0$ on the unphysical
real axis for both signs of $\epsilon.$
This pole stays away from the threshold even when $\epsilon\rightarrow 0$,
see Fig.~\ref{Fig:edge}. 

When the temperature is decreased to the region $\epsilon<0$,
$G^{-1}_\sigma$ has no roots on the physical real axis.
The pole, which did correspond to the stable particle moves continuously 
over the threshold to the unphysical real axis. 
In this temperature range this solution is denoted by $\delta_-$. This 
means that for
$\epsilon<0$ one has two poles $(\delta_-,\tilde\delta)$ 
on the unphysical real axis for sufficiently small~$|\epsilon|$.

In this approximation the spectral function 
$\rho(\epsilon,\delta)=-{\rm Im} G_\sigma(p_0)/\pi$ 
has the following parameterization above the threshold $p^2_0>M^2$:
\be
\rho(p_0,\epsilon)=-
\frac{\left[1+\frac{c}{b_1}\epsilon\right](b_2\pi)^{-1}
\sqrt{\frac{p_0^2}{M^2}-1}(\tilde\delta+\delta_\pm)}
{(\frac{p_0^2}{M^2}-1-\delta_\pm\tilde\delta)^2+(\frac{p_0^2}{M^2}-1)
(\tilde\delta +\delta_\pm)^2}
\label{spectr2}
\ee
(the expression depends on $\epsilon$ also indirectly, through the $\tilde
\delta\cdot \delta_\pm$ term).
The pole terms associated with $\delta_+$ and $\delta_-$ in the respective
temperature ranges $\epsilon>0$ and $\epsilon<0$  can be uniquely 
separated in $G_\sigma$. The approximate 
spectral function based exclusively on this singular ``pole'' 
contribution, above the threshold $M$ reads:
\be
\rho_{\rm sing}={(1+c\epsilon/b_1)\over b_2(\delta_\pm-\tilde\delta)
\pi}\times{|\delta|\over |\delta|^2+\delta_\pm^2}.
\label{appspectr}
\ee

 The maximal threshold
enhancement clearly occurs for $\epsilon =0$, when $\rho_{\rm sing}\sim
1/|\delta|$, which has been found also in Ref.~\cite{hatsu}. 
Since $\delta_+(\epsilon)$ and $-\delta_-(\epsilon)$ in the respective
temperature ranges are equal to linear order in $\epsilon$,  
the mirror variation to 
${\cal O}(\epsilon)$ of $\rho_{\rm sing}$ around $T^*$ is deformed only
by the temperature dependent first factor on the right hand side of 
Eq.~(\ref{appspectr}). 

The threshold enhancement sets in in a universal way, 
reflecting the motion of $\delta_+ (\delta_-)$ along the physical
(unphysical) real axis towards the position of the threshold. Concerning 
universality based on the expansion given in Eq.~(\ref{expand}) three further 
remarks are in order: i) the linear dependence of $\delta_\pm$ on 
$\epsilon$ is universal, which determines the leading term of the 
enhancement according to the second factor in Eq.~(\ref{appspectr}),
ii) when looking at the first correction to this leading 
behavior the one-pole description is not sufficient, $\tilde\delta$ plays
an important role (if the expansion is pushed to higher orders,
further roots would appear), iii)~it is obvious that the expansion given in Eq.~(\ref{expand})
is independent of the low temperature quasiparticle spectra.

Another characteristic temperature value  may be defined, corresponding to 
the temperature where $\delta_-$ and $\tilde\delta$ become degenerate 
on the unphysical real axis, when the temperature is gradually
decreased further. 
It is determined by the smaller negative root of the equation
$b_1^2-4b_2(a_1\epsilon_0+a_2\epsilon_0^2)=0.$
For smaller temperature values $(\epsilon<\epsilon_0)$ the two poles
become complex. One of the poles moves into the second Riemann sheet
 and can be interpreted as a resonance of finite lifetime if
the temperature becomes sufficiently low. The other
pole is its ``mirror'' located on the continuation of the second Riemann
sheet through the unphysical real axis into the positive halfplane, and 
does not have any 
physical meaning. This picture fully agrees with the qualitative pole
trajectory found in \cite{nagyN2}.
It could happen that the validity of the quadratic expansion of 
$G_\sigma^{-1}$ breaks down already before the degeneracy of the two
poles on the unphysical real axis occurs and $\delta_-$ may stay on this
axis until the temperature is decreased to zero. This alternative 
seems to be realized in the calculation presented in \cite{hidaka}.
It would be very interesting to see the presence of such a pole
in the low temperature representation of the pion-pion scattering 
amplitude based on chiral perturbation theory \cite{dobado}.

The extension of the investigation to the case of nonzero spatial momentum 
$|\p|/M$, regarded as a new small parameter can be readily done, by assuming
the analyticity  of $G_\sigma$ also on this quantity. Now the 
threshold parameter $\delta$ takes its original Lorentz invariant 
form: $\delta=\sqrt{1-p^2/M^2}$, where $p^2$ is the squared
four-momentum. At finite temperature, in the propagator 
also a separate dependence on the spatial momentum appears. Therefore
the expansion Eq.~(\ref{expand}) will contain 
also terms of the form $(\p^2/M^2)^l\epsilon^k\delta^m$ (note, that
$\epsilon$ remains as it was introduced for $|\p|=0$!). To
lowest order (single pole approximation) one keeps terms linear in
$\epsilon$ and $\delta$ and the coefficient $d$ of a new term 
$d\times|\p|^2/M^2$ has to be computed. 

The coefficient $d$ determines the shift in the value of the 
temperature for which the maximal threshold enhancement of
$\rho_{sing}$ occurs at small finite $|\p|$:
$T^*(|\p|)=T^*[1+\frac{d}{4|a_1|}\frac{\p^2}{m_\pi^2}]$. 
Since the term proportional to $\p^2$ is real even after the
continuation above the threshold, $|\p|$ and $\epsilon$ play similar
roles in the spectral function. For fixed $T$, it is $d$, which
controls the location and height of the maximum of $\rho_{sing}$ when 
$|\p|$ is increasing. For instance, one finds  
$\delta_{max}\approx |a_1\epsilon+d|\p|^2/M^2|/b_1$.
 
Below, we illustrate the validity of the simplified representation 
discussed above for  the scalar-isoscalar spectral function of 
the linear $\sigma$ model, obtained to leading order in an expansion 
with respect to the inverse of the number of the Goldstone bosons 
(large $N$ approximation). For this approximate solution one checks explicitly
the existence of the power series of $G^{-1}_\sigma$ 
in $\delta,$ $|\p|$ and $\epsilon$.
Furthermore, the ranges of validity in $|\epsilon|$ of the single
and the two pole descriptions of the spectral function can
be analyzed by comparing Eq.~(\ref{spectr2}) and
Eq.~(\ref{appspectr}) with the complete
expression of the spectral function in the model.

The first goal in the second part is to sketch the way the $\sigma$ propagator
given by Eq.~(19) of Ref.~\cite{nagyN2} can be cast for 
$p_0<M\equiv 2m_\pi(T)$ and $|\p|=0$
in the form presented in Eq.~(1) of the present paper. Employing
an obvious notation, the 
renormalized couplings $(m_R^2,\lambda_R, h)$
of this model were fixed at $T=0$ at values which ensure
$m_{\pi}(0) =140~{\rm MeV},~f_\pi=\sqrt{N}\Phi(T=0)=93~{\rm MeV},$
$m_\sigma=3.95 f_\pi,$
$m_\sigma/\Gamma_\sigma\sim 1.4$. 
The temperature dependence of the vacuum expectation value of the $O(N)$
field, $\Phi(T)$,  was determined by the equation of state 
(its renormalized form appears in Eq.~(11) of Ref.~\cite{nagyN2}):
 \be
m^2+\frac{\lambda}{6}\left[\Phi^2(T)+\int_k
\left[n(\omega_k)+\frac{1}{2}\right]\frac{1}{\omega_k}\right]-
\frac{h}{\Phi(T)}=0,
\label{eqstate}
\ee 
where $n$ is the Bose-Einstein factor and $\omega_k^2=\k^2~+~m_\pi^2$.
The expansion of $m_\pi(T)$ to quadratic order in $\epsilon$ is
provided through the combination of the relation $m_\pi^2(T)=h/\Phi(T)$ 
with the expansion of $\Phi(T)$ around $\Phi(T^*)$. 
$\Phi(T^*)$ is found from Eq.~(31) of  Ref.~\cite{nagyN2}. 

The selfenergy of $\sigma$ is determined to leading
large $N$ order by a sum of the chain of pion bubbles, characterized
by the function $b^{>}(p_0)$
 as shown by Eq.~(5) of Ref.~\cite{nagyN2} (the index $>$ refers to
the first Riemann sheet). 
The expression of the bubble above the threshold given in  
Ref.~\cite{nagyN2}
is written, after continuing to the physical real axis below the threshold 
and some convenient integral transformations, in a form where
the expansion in powers of $\delta$ and $\epsilon$ actually starts from:
\bea
\nonumber
4\pi^2b^>(p_0)=\frac{1}{4}\ln\frac{m_\pi^2(T)}{M_0^2}
+\frac{\delta\arccos(\delta)}{2\sqrt{1-\delta^2}}\\
-\int\limits_\xi^\infty dx
\frac{(x^2-\xi^2)^{-\frac{1}{2}}}{e^x-1}
+\delta\int\limits_0^\infty dx \frac{[(1+x^2)\sqrt{1+\delta^2x^2}]^{-1}}
{e^{\xi\sqrt{1+\delta^2x^2}}-1},
\label{b>_eq}
\eea
where $m_\pi(T)$ and $\xi =m_\pi(T)/T$ carry the explicit $T$ dependence 
($M_0$ is the normalization scale of the theory defined at $T=0$). 
One performs first the expansion in powers of $\delta$ to quadratic order.
For the second integral in Eq.~(\ref{b>_eq}) one finds to linear order
$(\frac{\pi}{2}-\delta)n(m_\pi(T))$.
Next, the remaining explicit pion mass dependence (also in $\xi$!) is 
expanded around $T^*$ with the procedure sketched above. 
 With these hints one easily establishes the
approximate expression Eq.~(\ref{expand}), obtaining numerically
the following
values for the coefficients:
$a_1=-18.32$, $a_2=-23.14$, $b_1=8.59$, $b_2=14.48$, $c=9.67$.

\begin{figure}[htpb]
\begin{center}
\includegraphics[width=7cm]{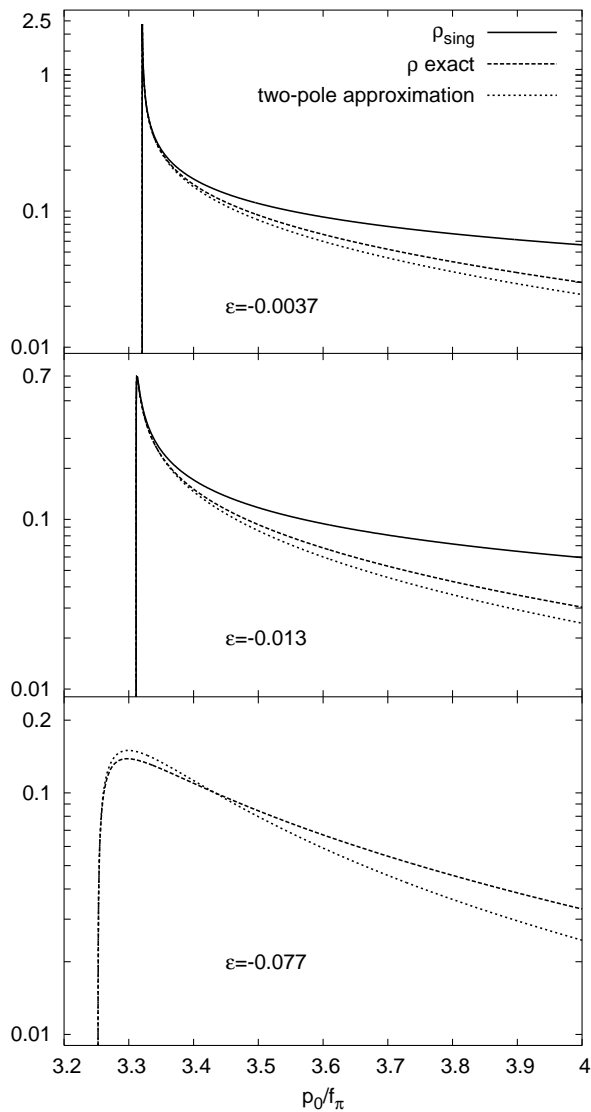}
\end{center}
\caption{
The spectral function $\rho(p_0,\epsilon)f_\pi^2$ in various approximations
as a function of $p_0$ at different reduced temperature $\epsilon$, and
$|\p|=0$.}
\label{spct}
\end{figure}

In Fig.~\ref{spct} the spectral function and its universal approximations are 
shown for three different temperatures. The curves labeled ``exact'' are
calculated from the complete expression derived in Ref.~\cite{nagyN2}. They
are compared with the two approximate forms described above.
All temperatures were chosen below $T^*$ (above it the situation is
qualitatively similar). For the highest value of $\epsilon$
the pole $\delta_-$ lies almost on the threshold, for its smallest value 
 $(\epsilon =\epsilon_0\approx-0.077)$ one has $\delta_-=\tilde\delta$.
When calculated from the exact expression one obtains: $\epsilon_0=-0.13.$
According to these curves the  region where the one-pole description gives 
 a good approximation is rather narrow in $\epsilon$ and in particular in 
$\delta.$ The correction due to
the second pole contribution is quite important for increasing
frequency values. For $\epsilon =\epsilon_0$ the line corresponding to 
$\rho_{sing}$ was omitted, since its  expression becomes meaningless.
It is, however, remarkable how good
approximation can be constructed using Eq.~(\ref{expand}). 
The range of validity of the quadratic expansion
of $G^{-1}_\sigma$ in $\epsilon$ 
includes also the degeneracy temperature of the poles,
persisting even for somewhat lower temperatures when both poles are complex 
and lie equidistantly from the threshold.
The specific form of the spectral function,
$\rho\sim|\delta|/(|\delta|^2+const)^2$, characteristic
at the degeneracy temperature is expected to occur even in the 
exact solution of the $N=4$ linear $\sigma$ model with 
physically relevant couplings.

For the case of the $\sigma-\pi$ system, the coefficient $d$, which 
determines the influence of the explicit $|\p|$-dependence,
  comes uniquely from the temperature dependent part of the pion bubble,  
$b_T^>\equiv b_T^>(p_0,|\p|)$.
 Since the $T=0$ part of the expression of the bubble
is Lorentz invariant, it remains unchanged when expressed with the
complete expression of
$\delta$. The expression of $b_T^>$ for $\p\ne 0$, see e.g. 
Refs.~\cite{hidaka,toni} can be cast in the form
\be
b_T^>=\int\limits_1^\infty \frac{dt}{8\pi^2} 
\frac{n(t m_\pi)}{|\p|}
\ln\Bigg|\frac{\left[\frac{p^2}{2}+|\p|\sqrt{t^2-1}
\right]^2-p_0^2t^2}{\left[\frac{p^2}{2}-|\p|\sqrt{t^2-1}
\right]^2-p_0^2t^2}\Bigg|.
\ee   
For the threshold behavior of the spectral function
 to the lowest (linear) order in $(\epsilon, \delta, \p^2)$ , 
one can use $\delta=0$ in the
integral, that is for the quantities scaled by $m_\pi(T)$ one can put: 
$p^2=4,$ $p_0^2=4+\p^2.$
With a suitable transformation of the variable of integration
we can  ensure the smoothness of the integrand, obtaining:
\be
b_T^>=\int\limits_0^\infty \frac{dt}{4\pi^2 \xi}
\ln\frac{1-\exp\big[-\xi
\sqrt{1+\big(\frac{|\p|}{2}\tanh\frac{|\p|}{2}t\big)^2}\big]}
{1-\exp\big[-\xi\sqrt{1+\big(\frac{|\p|}{2}\coth\frac{|\p|}{2}t\big)^2}
\big]}.
\ee
Performing a partial integration one obtains the first two terms of
the asymptotic expansion of the integral valid for small $|\p|$, relevant
for our analysis. Next to the unchanged $|\p|$-independent term 
with  the complete $\delta$ one can calculate the coefficient of $\p^2$ 
by performing the $t$-integral with the value of $\xi$ determined at $T^*$. 
For the actual couplings $d=1.65$ is found.

The assumed expansion of the inverse propagator around the threshold in
powers of $\epsilon$ and $\delta$ resembles the starting point of the Landau
theory of second order phase transitions. Like in the Landau theory, our
purpose has not been to prove the analytic property of the relevant
quantity, but to explore the consequences of its postulation.
We conjecture the universality of the threshold behavior in
any channel where a particle pole describing at high temperature a stable
particle might pass through the threshold of its lowest two-body decay. Our
discussion implies that for
$T\lesssim T^*$ the pole dominating analytically the spectral function
cannot correspond to any particle. Whether this remains valid down to 
$T=0$ is model dependent. 

A possible generalization of Eq.~(\ref{spectr2}),
in which only the quadratic dependence of the inverse propagator on 
$\delta$ is assumed,  can be easily written down:
\be
\rho (\epsilon,\p,p^2)=\frac{R_1(\epsilon,\p)\sqrt{\frac{p^2}{M^2}-1}}
{[\frac{p^2}{M^2}-1+R_2(\epsilon,\p)]^2+R_3(\epsilon,\p)[\frac{p^2}{M^2}-1]}.
\ee
The functions $R_i$ are real, at $T=0$ they assume constant values.
$\epsilon$ is measured from the value $T^*$ determined by
the condition $R_2(\epsilon =0,0)=0$. A sequence of reduced temperatures
 $\epsilon(\p)\equiv (T^*(\p)-T^*)/
T^*$ can be defined as the solution of the equation 
$R_2(\epsilon(\p),\p)=0$. Obviously $T^*(0)=T^*$. Asymptotically
 one expects the ``scaling'' behavior
$R_2\sim [\epsilon-\epsilon(\p)]^\alpha,\alpha>0.$
The treatment based on Eq.~(\ref{expand}) leads to $\alpha =1$, but in
principle a nontrivial non-integer scaling exponent can be envisaged.
Note, that when there is no explicit symmetry breaking the branch point 
tends to the origin if $|\p|$ goes to zero 
and $T^*=T_c$ independently of $|\p|$ \cite{nagyN1}.

The next step will be to discuss the effect of the finite pion lifetime.
It shifts the branch-point into the lower halfplane.
Around it there is still an enhancement (similar to the undamped case) but
in the spectral function varying along the real axis the effect is smeared
out partially, see \cite{ohtani}. Either a complete two-loop (improved)
perturbative analysis or the evaluation of the next-to-leading large N
correction to $G_\sigma$ is needed to reliably assess the importance
of the finite pion width.

This research has been supported by the research contract
OTKA-T037689 of the Hungarian Research Fund.

\end{document}